\documentclass[reprint,prb,final,superscriptaddress,longbibliography,nobibnotes,footinbib,amsfonts,showpacs,letterpaper,amssymb,pdftex]{revtex4-1}
\usepackage{graphicx,multirow,bm,amsmath}
\usepackage{booktabs}
\usepackage{hyperref}
\usepackage{bbm}
\usepackage[usenames,dvipsnames]{color}
\usepackage[table,xcdraw]{xcolor}

\newcommand{\kbrnasaames}{KBR, Inc., Intelligent Systems Division, NASA Ames Research Center, Moffett Field, CA 94035, USA}

\begin{document}

	\title{Globally Optimal Band Structure for Thermoelectrics in Realistic Systems}
	\author{Junsoo Park}
	\email{junsoo.park@nasa.gov}
	\affiliation{\kbrnasaames}
	\date{\today} 
%	\pacs{******}
	\begin{abstract}
Observation is made that a linear dispersion in any dimension under acoustic-phonon-deformation-potential scattering theoretically prescribes a constant charge transport distribution, required for the boxcar profile known to maximize the thermoelectric figure of merit. A linear dispersion squeezed by two transport gaps for optimized bandwidth under scattering by phonon deformation then theoretically constitutes a globally optimal qualitative band structure that may arise in real systems.
	\end{abstract}
	\maketitle

\section{Introduction}

\begin{figure*}[tp]\centering
\includegraphics[width=0.9 \linewidth]{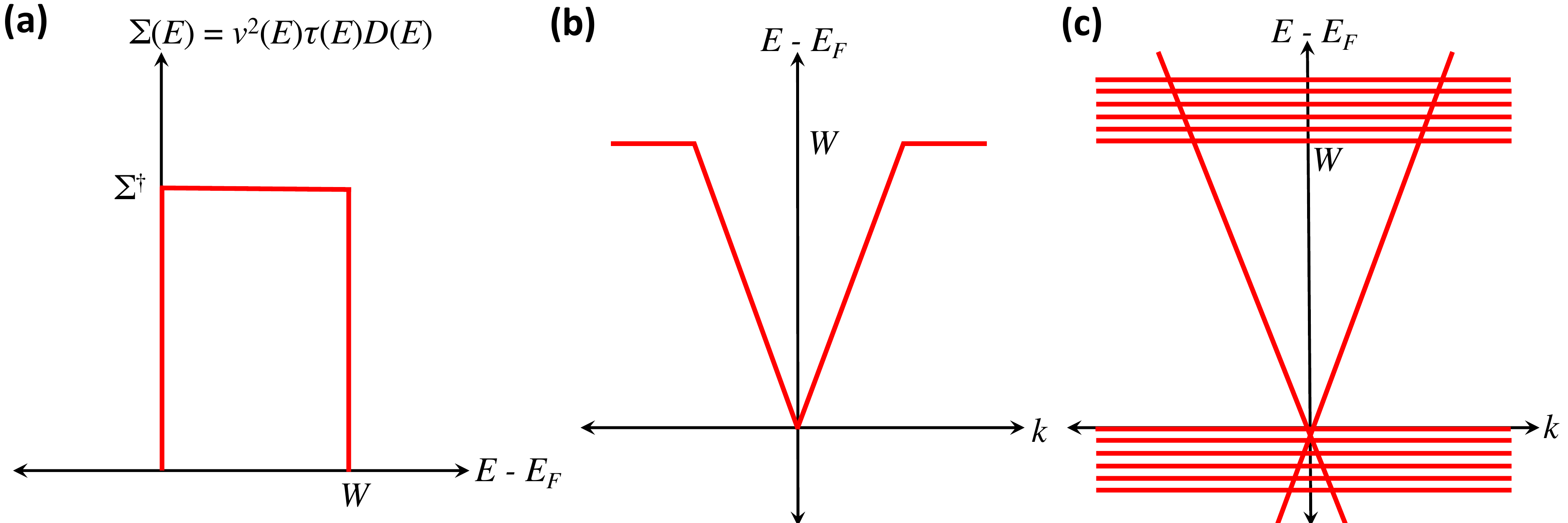}
\caption{ \textbf{a)} A boxcar transport distribution. \textbf{b)} Schematic of the ideal band structure achieving a boxcar transport distribution with a band gap and high-energy cutoff. \textbf{c)} Semimetallic schematic where transport cutoffs are achieved by resonant scattering states. All diagrams are from the $n$-type perspective where electron dominates transport as opposed to hole. The energy axes are zero-referenced to the Fermi level for the sake of graphics, which in practice is an optimization parameter.}
\label{fig:main}
\end{figure*}

In electron band transport, the charge transport distribution $\Sigma(E)$ (a.k.a. spectral conductivity, energy-dependent conductivity, transmission function) determines key properties such as Ohmic conductivity, the Seebeck coefficient, and electronic thermal conductivity. It is known that a boxcar-shaped $\Sigma(E)$ for given magnitude delivers optimal thermoelectric figure of merit $zT$ in quantum systems obeying the Landauer-B{\"u}ttiker transport formalism \cite{mostefficientquantumthermoelectric,mostefficientquantumthermoelectric2} as well as the semiclassical Boltzmann transport equations (BTE) under the relaxation time approximation \cite{searchforthebestthermoelectric,limitsofthermoelectricperformance}. The latter has recently been derived by exact mathematics \cite{limitsofthermoelectricperformance}. Under BTE, $\Sigma_{x}(E) = \frac{1}{V} \langle v_{x}^{2}(E)\rangle D(E)\tau(E)$, where $V$ is the system volume, $x$ is a Cartesian direction, $\langle v_{x}^{2} \rangle$ is the squared group velocity averaged over constant energy surface, $D$ is the density of states, and $\tau$ is the lifetime, or inverse of the scattering rate $\tau^{-1}$. A boxcar function, shown in Fig. \ref{fig:main}a, is essentially a Heaviside step function with an upper cutoff just like the lower one. Within their confines, the profile is constant and flat, or $\Sigma(E)=\Sigma^{\dagger}\propto E^{0}$. The ultimate question yet to be answered is what, if any, realistic bulk band structures and systems may physically achieve this and deliver optimal thermoelectric performance. It is herein observed that linear dispersion, in any dimension, under scattering by acoustic phonon deformation meets the theoretical requirement for realizing this goal. 

\section{Linear Dispersion Under Deformation-Potential Scattering}

Table 1 summarizes the $E$-dependence of $\langle v_{x}^{2}(E) \rangle$ and $D(E)$ of isotropic parabolic and linear dispersions in one, two, and three dimensions, along with the required $E$-dependence of $\tau(E)$ and the corresponding $D(E)$-dependence of $\tau^{-1}(E)$ to enforce $\Sigma(E)\propto E^{0}$. No parabolic case is realistic, as they require $\tau^{-1}(E)$ profiles that do not match, even approximately, with any known electron scattering mechanism. In turn, all of the linear cases, which exhibit $\langle v_{x}^{2}(E) \rangle \propto E^{0}$, require that $\tau^{-1}(E) \propto D^{1}(E)$, the very behavior of deformation-potential scattering by long-wavelength longitudinal acoustic phonons - this has been derived not only for spherical Fermi surfaces of parabolic bands specifically \cite{seitz,lundstrombook,rodebook,nagbook} but also for an arbitrary Fermi surface around some band minimum \cite{radcliffe,ziman}. More generally, consider a $d$-dimensional dispersion of order $p$. It is straightforward to show $\langle v^{2}(E) \rangle \propto E^{\frac{2(p-1)}{p}}$ and $D(E)\propto E^{\frac{d}{p}-1}$. The scattering behavior required to enforce $\Sigma(E)\propto E^{0}$ then is $\tau^{-1}(E) \propto E^{\frac{p+d-2}{p}}$, or in a more useful form, $\tau^{-1}(E) \propto D(E) E^{\frac{2(p-1)}{p}}$. The only realistic solution is $p=1$, or linear dispersion, regardless of dimension under acoustic-deformation-potential scattering. This solution is as consequential as it is simple because linear dispersion is common and acoustic-deformation-potential scattering is ubiquitous in real materials.

\begin{table}[]\label{table}
\begin{tabular}{cccccc}
\multicolumn{1}{c|}{Parabolic}       & $\langle v_{x}^{2}(E) \rangle$ &     $D(E)$         &   $\tau(E)$        & $\tau^{-1}(E)$   & $\Sigma_{x}(E)$ \\ \hline
\multicolumn{1}{c|}{1D}      &    & $E^{-\frac{1}{2}}$  & $E^{-\frac{1}{2}}$       & $D^{-1}(E)$      &    \\ 
\multicolumn{1}{c|}{2D}      & $E^{1}$    & $E^{0}$                 & $E^{-1}$                     & $D^{\infty}(E)$   &   $E^{0}$     \\
\multicolumn{1}{c|}{3D}      &     & $E^{\frac{1}{2}}$   & $E^{-\frac{3}{2}}$       & $D^{3}(E)$      &   \\  \cline{1-2} \cline{2-6} \\
\multicolumn{1}{c|}{Linear}               & $\langle v_{x}^{2}(E) \rangle$ & $D(E)$     & $\tau(E)$  & $\tau^{-1}(E)$  & $\Sigma_{x}(E)$ \\ \hline
\multicolumn{1}{c|}{1D}       &       & $E^{0}$    & $E^{0}$  & $D^{1}(E)$  &   \\
\multicolumn{1}{c|}{2D}       & $E^{0}$    & $E^{1}$    & $E^{-1}$ & $D^{1}(E)$   &   $E^{0}$     \\
\multicolumn{1}{c|}{3D}       &       & $E^{2}$    & $E^{-2}$ & $D^{1}(E)$  & \\ \cline{1-2} \cline{2-6}
\end{tabular}
\caption{Energy-dependence of intrinsic isotropic band structure features as well as that of lifetimes and the causal carrier scattering rates required for enforcing $\Sigma(E)\propto E^{0}$.}
\end{table}

To ascertain the need for deformation-potential scattering, consider the generalized deformation potential derived by Kahn and Allen \cite{defpotkahnallen}, whose tensor form is
\begin{equation}\label{eq:kahnallen}
\mathbf{\Delta}_{\mathbf{k}} = \mathbf{\Delta}_{\mathbf{k}}^{0}+\mathbf{v}_{\mathbf{k}}\otimes\mathbf{v}_{\mathbf{k}}
\end{equation}
where $\mathbf{v}$ is the group velocity vector, and  $\Delta^{0}$ the usual Bardeen-Shockley term \cite{defpotbardeenschokley} representing strain-driven energy shift of bands. Note that the second, correction term does not contribute any energy-dependence for linear bands since $\langle\mathbf{v}\rangle$ is a constant of energy. Since $\Delta$ is essentially energy-independent for a single band, the scattering rate is guaranteed to behave as $\tau^{-1}(E)\propto D^{1}(E)$ due to the near elasticity of the process. Eq. \ref{eq:kahnallen} has been implemented in the AMSET software \cite{amset} and has led to very good agreements with experimentally measured transport properties for numerous materials. Though optical and high-wavevector phonon deformations can introduce some inelasticity in real systems, first-principles calculations of full electron-phonon scattering in three \cite{bandconvergencenotbeneficial,analoguepbte,cosiyi} and two \cite{grapheneepw1,mos2yi} dimensions, as well as other analyses \cite{positiveseebeckli}, have demonstrated that the $\tau^{-1}(E)\propto D^{1}(E)$ trend persists for scattering due to phonon deformation, as phonon energies are typically small in the electronic energy-scale. Other common processes, such as polar-optical or ionized-impurity scattering in semiconductors, are not known to exhibit as consistently simple $E$-or-$D(E)$-dependence of rates. The former has discontinuities at polar optical phonon energies with different pre-emission and post-emission behaviors, while the latter has a logarithmic screening term \cite{ziman,lundstrombook,ionizedimpurity}. As such, they cannot be expected to pair with a simple realistic band to robustly yield $\Sigma(E)\propto E^{0}$.

\section{Optimum Bandwidth and \textit{\lowercase{z}T}}

The $\Sigma(E)\propto E^{0}$ profile is of course only one of the three requirements for realizing a boxcar function. The second requirement is a lower cutoff, and ideally a band gap would guarantee it as in Fig. \ref{fig:main}b. Unfortunately, a true linear dispersion would be gapless, forming a Dirac cone at the Fermi level with symmetric dispersion on the other side, triggering bipolar transport that suppresses thermopower. The dispersion would therefore need to develop a finite curvature at its utmost tip, similar to a Kane band \cite{kane1,kane2} with a huge non-parabolicity parameter. In this semiconducting case, for polar-optical scattering to be a non-factor, the material would preferably be non-polar. If polar, its dielectric constant must have tiny ionic part, or the operation temperature must be below that seriously activates polar phonons. If the degeneracy at the Dirac point cannot be lifted, then the viable picture is for the opposing carrier type to be suppressed in $\tau$ to near-zero values by scattering into heavy resonance-like states, as in Fig. \ref{fig:main}c. This concept has been demonstrated in real semimetals with linear dispersion with heavy-band crossings \cite{cosiyi,mos2yi}, if not quite so ideally as here because the heavy states are not so flat and inelastic scattering slightly broadens the resolution of the boxcar edge. Nevertheless, considering the generally low phonon energies ($\lesssim 30$ meV for decent thermoelectrics) in the energy scale of electronic bands ($\lesssim 1$ eV), only the states close to the resonance states would be affected. The lower cutoff alone creates a Heaviside $\Sigma(E)$, which would be optimal for the power factor but not $zT$ \cite{limitsofthermoelectricperformance}.

\begin{figure*}[tp]\centering
\includegraphics[width=1 \linewidth]{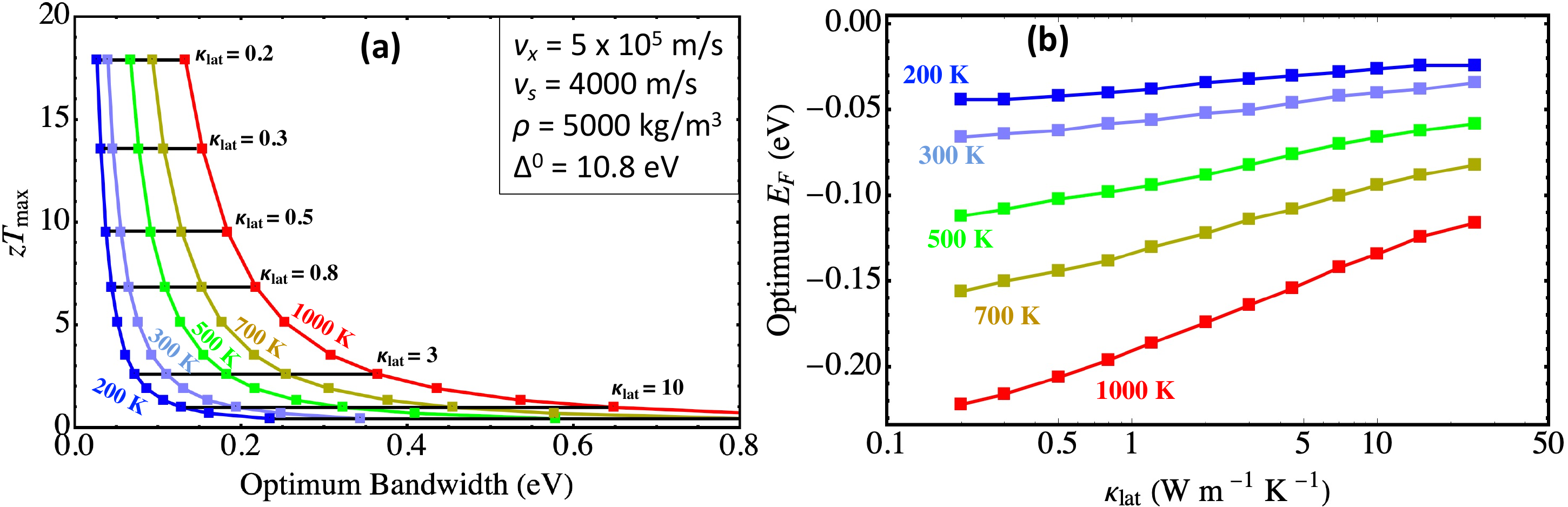}
\caption{ \textbf{a)} $T$-and-$\kappa_{\text{lat}}$-dependent optimum bandwidth and $zT$ under deformation-potential scattering for an isotropic 3D linear band with $v_{x}=5\times10^{5}$ m/s, which is the group velocity of a parabolic band with effective mass 0.067 (approximately that of GaAs) 0.1 eV above the band edge. All other material parameters are kept identically to the analysis in Ref. \onlinecite{optimalbandstructure}. The $\kappa_{\text{lat}}$ values are given in W m$^{-1}$ K$^{-1}$. \textbf{b)} Optimum Fermi level (electron chemical potential) for each case.}
\label{fig:bandwidth}
\end{figure*}

The third and final requirement for completing the boxcar is an upper cutoff. The reason for this is that, whereas the power factor does not concern electronic thermal conductivity ($\kappa_{e}$), $zT$ does, and high-energy states contribute more to thermal conduction than charge conduction compared to low-energy states in relative terms. Shutting down transport of high-energy carriers is especially relevant when $\kappa_{e}>\kappa_{\text{lat}}$, that is when $\kappa_{\text{lat}}$ is low and $\kappa_{e}$ and the PF are high, precisely the condition for high performance. For this to happen, there would ideally exist another energy gap at $W$ with the linear dispersion discontinuously flattening out, as in Fig. \ref{fig:main}b, though realistically the flattening would be abrupt at best. Perhaps a more viable picture is for efficient scattering states to be present or introduced so that $\tau$ is essentially zeroed out above $W$ mirroring the low-energy resonance states, as in Fig. \ref{fig:main}c. In either case, the resulting bandwidth $W$ is an important optimization parameter as investigated in Ref. \onlinecite{optimalbandstructure} and is narrower at lower lattice thermal conductivity ($\kappa_{\text{lat}}$) and higher velocity. The theoretical optimum bandwidth ($W_{\text{opt}}$) and the corresponding maximal $zT$ for linear dispersion under deformation-potential scattering are shown in Fig. \ref{fig:bandwidth}. $W_{\text{opt}}$ is obtained by the approach used in Ref. \onlinecite{optimalbandstructure}, solving for $W_{\text{opt}} = \text{argmax}_{W} zT(W)$ where
\begin{widetext}
\begin{equation}\label{eq:ztnew}
zT(W) = \frac{ \left( \int_{0}^{W} (E_{\text{F}}-E)\Sigma(E) \left(-\frac{\partial f}{\partial E}\right)dE \right)^{2} }{ \int_{0}^{W} \Sigma(E) \left(-\frac{\partial f}{\partial E}\right)dE \left[ T\kappa_{\text{lat}} + \int_{0}^{W} (E_{\text{F}}-E)^{2}\Sigma(E) \left(-\frac{\partial f}{\partial E}\right)dE \right] - \left( \int_{0}^{W} (E_{\text{F}}-E)\Sigma(E) \left(-\frac{\partial f}{\partial E}\right)dE \right)^{2}},
\end{equation}
\end{widetext}
and $\tau$ is calculated using Eq. \ref{eq:kahnallen} and 
\begin{equation}\label{eq:defpottau}
\tau(E)=\frac{\rho v_{s}^{2}}{\pi k_{\text{B}}T}\frac{D^{-1}(E)}{(\Delta^{0}+\langle\mathbf{v}\cdot\mathbf{v}\rangle)^2},
\end{equation}
where $v_{s}$ is the sound velocity, $\rho$ is the mass density, and $\langle\mathbf{v}\cdot\mathbf{v}\rangle=\langle v_{x}^{2}(E)\rangle=v_{x}^{2}$ for an isotropic linear band. For the sake of comparison with previous results for a parabolic band \cite{optimalbandstructure}, the same material parameters are maintained. Three noteworthy observations are made here.

1) Linear dispersion prescribed with optimized width indeed outperforms parabolic dispersion of comparable velocity and optimized width. The improvement is particularly pronounced at low temperatures. At 200 K, linear dispersion with $W_{\text{opt}}$ yields nearly five times higher $zT$ than the parabolic counterpart. In fact, $zT_{\text{max}}$ is identical for all temperatures for a given $\kappa_{\text{lat}}$ value and is in excellent agreement with the analytically derived results of Ref. \onlinecite{limitsofthermoelectricperformance} purely from boxcar $\Sigma(E)$. For $\kappa_{\text{lat}}=0.2$ W m$^{-1}$ K$^{-1}$, we have $\frac{\kappa_{\text{lat}}}{k_{\text{B}}^{2}T\Sigma^{\dagger}}\approx0.0209$ which is temperature-independent due to the $T^{-1}$ factor in $\Sigma^{\dagger}$ through Eq. \ref{eq:defpottau}. This quantity converts to $zT_{\text{max}}\approx18$ both analytically and by the present calculation. This is the highlight for truly optimized band structures: the electronic performance is temperature-independent, and the temperature-dependence of $zT$ owes solely to that of $\kappa_{\text{lat}}$. A band's linearity and its bandwidth optimization are thus critical for low temperatures notoriously barren of high $zT$.

2) $W_{\text{opt}}$ for a linear band is roughly half that of a parabolic band with a similar velocity. This is ascribed to the $\Sigma(E)\propto E^{0}$ trend, which is much less needy of high-energy states to drive thermopower than the $\Sigma(E)\propto E^{1}$ trend of a parabolic band. It thus becomes more beneficial to trim high-energy contributions, sacrificing a bit of thermopower in exchange for reducing electronic thermal conductivity, which naturally narrows $W_{\text{opt}}$ with appropriate adjustment in $E_{\text{F}}$.

3) For an isotropic linear dispersion under Eq. \ref{eq:defpottau},
\begin{equation}\label{eq:sigmalimit}
\Sigma^{\dagger} = \frac{\rho v_{s}^{2}}{V \pi k_{\text{B}}T}\frac{v_{x}^{2}}{(\Delta^{0}+v_{x}^{2})^{2}},
\end{equation}
meaning there is an optimum velocity for maximizing $\Sigma^{\dagger}$ and by default $zT$, precisely at $v^{2}=\Delta^{0}$, or $v=\sqrt{\Delta^{0}}$. For $\Delta^{0}=10$ eV, this corresponds to $v\approx1.32\times10^{6}$ m/s which turns out to be close to the Fermi velocity of graphene. Any higher velocity will, insofar as the Kahn-Allen potential is valid, in fact lower $zT$. At 300 K, $W_{\text{opt}}$ for this velocity with $\kappa_{\text{lat}}=0.2$ W m$^{-1}$ K$^{-1}$ is approximately 30 meV in theory, though realistically closer to 50 meV if not higher depending on the maximum optical phonon energy whose deformation causes inelastic scattering. It translates to $zT_{\text{max}}\approx32$.

\begin{figure*}[tp]\centering
\includegraphics[width=1 \linewidth]{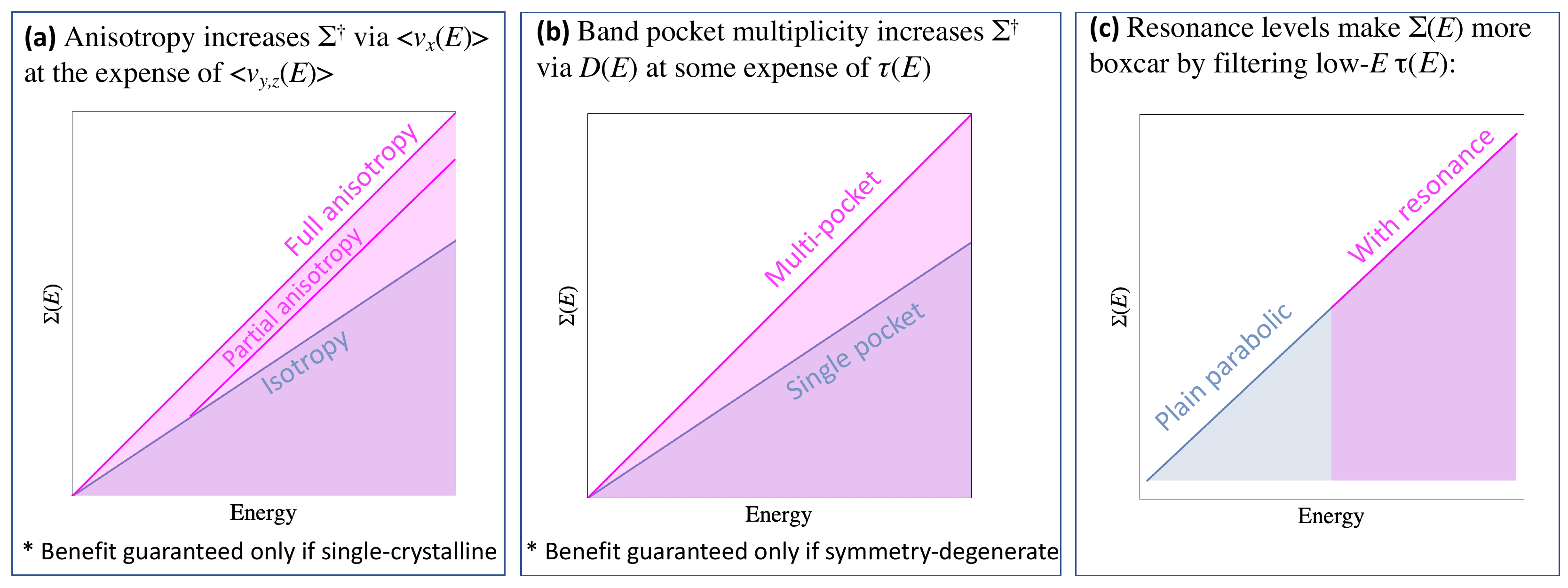}
\caption{Schematics of the effects of \textbf{a)} band anisotropy, \textbf{b)} band multiplicity, and \textbf{c)} resonance levels on the $\Sigma(E)$ profile of a parabolic band, whose basic trend under deformation-potential scattering is $\Sigma_{x}(E)=\Sigma^{\dagger}E$ due to $\langle v_{x}^{2}(E) \rangle \propto E$. Notice that their benefits come by either increasing $\Sigma^{\dagger}$ or making $\Sigma(E)$ more boxcar in shape.}
\label{fig:others}
\end{figure*}

\section{Discussions}

Of note, Ref. \onlinecite{bestthermoelectric} derives that $\Sigma(E)\propto\delta$ is optimal. This is in fact precisely the limit $W_{\text{opt}}\rightarrow0$ as either $\kappa_{\text{lat}}\rightarrow0$ or $v\rightarrow\infty$. However, $\kappa_{\text{lat}}\rightarrow0$ is unrealistic in real solids, and $v\rightarrow\infty$ is at odds with the band turning completely flat such that $\Sigma(E)\propto\delta$ is possible. In fact, $v\rightarrow0$ for a flat band, yielding zero conductivity and zero power, failing to deliver non-zero $zT$ unless in the unrealistic case of $\kappa_{\text{lat}}\rightarrow0$. In theory, even if $\kappa_{\text{lat}}\rightarrow0$ were possible in which case $zT\rightarrow\infty$ and the Carnot efficiency is reached, it would only represent an extremely slow, reversible process that virtually does not occur, to no engineering relevance. These have been pointed out by past works \cite{mostefficientquantumthermoelectric,mostefficientquantumthermoelectric2,limitsofthermoelectricperformance,optimalbandwidth,optimalbandstructure}. In practice, even a completely localized charge may be able to move via small-polaronic activated hopping, as occurs in molecular crystals \cite{chargetransportmolecular}, allowing for non-zero conductivity, power, and $zT$ alike. Unfortunately, the mobility resulting from this mechanism is generally so low ($\mu\lesssim10^{-2}$ cm$^{2}$ V$^{-1}$ s$^{-1}$) that conductivity of at most $\sigma\approx10^{3}$ S m$^{-1}$ can be expected at a very high, metallic carrier concentration of $10^{22}$ cm$^{-3}$. This conductivity is so low that, even when paired with a very high Seebeck coefficient of 300 $\mu$V K$^{-1}$ and the lowest practically imaginable value of $\kappa_{\text{lat}}\approx0.1$ W m$^{-1}$ K$^{-1}$, the resulting $zT$ is less than 0.5 at 300 K. This is clearly far from optimal. Therefore, the bandwidth must conclusively be finite and optimized, with the transport mechanism remaining high-velocity band-motion. 

Further, the linear-band carriers within $W_{\text{opt}}$ must not inelastically scatter into the heavy states at the cutoffs, which would drastically reduce $\tau$ and $\Sigma^{\dagger}$ everywhere, thereby $zT$. This means that, given a material-dependent maximum phonon energy $\omega_{\text{max}}$ that couple inelastic scattering, which is typically on the order of 10 meV, in practice one needs $W_{\text{opt}}^{*}=W_{\text{opt}}+\omega_{\text{max}}$ if heavy states exist on only one side (Fig. \ref{fig:main}b) and $W_{\text{opt}}^{*}=W_{\text{opt}}+2\omega_{\text{max}}$ if heavy states exist on both sides (Fig. \ref{fig:main}c). These would allow preservation of constant $\Sigma(E)$ with width $W_{\text{opt}}$, except with the boxcar edges broadening and slanting due to the inelastic effects. The whole profile would be rendered somewhat trapezoidal. To minimize the effect of inelasticity in practice, $\omega_{\text{max}}$ must be as small as possible, i.e. the material must be as soft as possible.

The present viewpoint also unifies the mechanisms by which other band structure features improve $zT$, namely band anisotropy, band multiplicity and resonance levels, whose salient effects have been systematically investigated in Refs. \onlinecite{optimalbandstructure,bandconvergencenotbeneficial}. In short, they all improve $zT$ via either making $\Sigma(E)$ more boxcar in shape or increasing its magnitude $\Sigma^{\dagger}$. Fig. \ref{fig:others} graphically summarizes each case for a parabolic band as an example, which has a $\Sigma(E)=\Sigma^{\dagger}E$ profile (linear in $E$) under deformation-potential scattering due to $\langle v^{2}_{x}(E)\rangle \propto E$. 

As Fig. \ref{fig:others}a shows, band anisotropy steepens $\Sigma(E)$, i.e., increases $\Sigma^{\dagger}$ by steepening $\langle v^{2}_{x}(E) \rangle$ due to the low-dimensional effect, though the location of steepening depends on the degree of anisotropy \cite{optimalbandstructure}. It also increases $D(E)$, but this is largely if not entirely cancelled by reduction in $\tau(E)$ under deformation-potential scattering. 

As Fig. \ref{fig:others}b shows, band multiplicity also steepens $\Sigma(E)$, i.e. increases $\Sigma^{\dagger}$, but through the enhancement of $D(E)$ relative to the reduction in $\tau(E)$ due to interband/intervalley scattering. Because scattering between bands/valleys is generally weaker than that within a band/valley, particularly if located at distant points in the \textbf{k}-space \cite{optimalbandstructure,bandconvergencenotbeneficial}, scattering does not increase as much as $D(E)$ does in the presence of multiple band pockets. For instance, if $D(E)$ increased by a factor of 2, $\tau(E)$ would decrease but less than by half, leading to an overall increase in $\Sigma^{\dagger}$ by a factor less than 2. Regardless, the overall $\tau^{-1}(E)\propto D(E)$ and $\Sigma(E)\propto \Sigma^{\dagger}E$ profiles are expected to persist under deformation-potential scattering.

Finally as Fig. \ref{fig:others}c shows, resonance levels benefit $zT$ by rendering $\Sigma(E)$ more boxcar in shape. By filtering out low-energy carriers via resonance scattering, it forms a steep edge in the $\Sigma(E)$ profile. Macroscopically, this has the effect of reducing the Ohmic conductivity relative to the thermoelectric conductivity, thereby enhancing the Seebeck coefficient. In fact, this is precisely the strategy to prescribe the boxcar edges depicted in Fig. \ref{fig:main}c, not only at low energies but also at high energies in order to curtail $\kappa_{e}$. Of course, the ultimate ingredient to perfecting the boxcar is a flat $\Sigma$-profile in between the edges, which is delivered by linear bands under scattering by phonon deformation, coming back full circle.

\section{Conclusion}

In summary, Fig. \ref{fig:main}b depicts a band structure shape that would generate a boxcar transport distribution under scattering by (acoustic) phonon deformation. Likelihood of achievement put aside, it is a qualitative optimal limit of a band structure in a realistic system that is physically conceivable and potentially closely emulated. Fig. \ref{fig:main}c depicts an alternative such design suiting semimetals where linear dispersion is common and deformation-potential scattering is nearly always dominant. Realizing the narrow optimum bandwidth would likely persist as the greatest challenge. Even if no system realizes it as far as to the ideal limit, known thermoelectric materials stand to potentially benefit from engineering the transport distribution such that it becomes more boxcar-like with little to no expense in $\Sigma^{\dagger}$. This study completes the study of Ref. \onlinecite{optimalbandstructure} that identified multiple band structure features that optimize $zT$ with the exception of its qualitative shape. Given now the ideality of linear dispersion under deformation-potential scattering, one with velocity approaching $\sqrt{\Delta^{0}}$ and symmetry-degeneracies (more carriers and/or less scattering) would lead to higher $\Sigma^{\dagger}$, while well-tuned bandwidth and Fermi level would optimize the Boltzmann transport integrals, en route to systematically and globally maximizing $zT$, to the benefit in particular of low-temperature performance integral to spacecraft propulsion and refrigeration.

\bibliography{references}

\end{document}